# Strong optical coupling in metallo-dielectric hybrid metasurfaces

**AJITH P RAVISHANKAR,**[1*] **FELIX VENNBERG,**[1] **AND ANAND SRINIVASAN**[1]

[1]*Department of Applied Physics, School of Engineering Sciences, KTH Royal Institute of Technology, Albanova University Center, Roslagstullsbacken 21, SE-106 91 Stockholm, Sweden*
*\*Corresponding author: aprav@kth.se*

Metasurfaces consisting of hybrid metal/dielectric nanostructures carry advantages of both material platforms. The hybrid structures can not only tightly confine subwavelength electromagnetic fields but also lower the absorption losses. Such optical characteristics are difficult to realize in metamaterials with only metal or dielectric structures. Hybrid designs also expand the scope of material choices and the types of optical modes that can be excited in a metasurface, thereby allowing novel light matter interactions. Here we present a metallo-dielectric hybrid metasurface design consisting of a high-index dielectric (silicon) nanodisk array on top of a metal layer (aluminium) separated by a buffer oxide (silica) layer. The dimensions of the nanodisks are tuned to support anapole states and the period of the nanodisk array is designed to excite surface plasmon polariton (SPP) at the metal-buffer oxide interface. The physical dimensions of the Si nanodisk and the array periods are optimized to excite the anapole and the SPP at normal incidence of light in the visible-NIR (400-900 nm) wavelength range. Finite difference time domain (FDTD) simulations show that, when the nanodisk grating is placed at a specific height (~200 nm) from the metal surface, the two modes strongly couple at zero detuning of the resonances. The strong coupling is evident from the avoided crossing of the modes observed in the reflectance spectra and in the spectral profile of light absorption inside the Si nanodisk. A vacuum Rabi splitting of up to ~ 129 meV is achievable by optimizing the diameters of Si nanodisk and the nanodisk array grating period. The proposed metasurface design is promising to realize open cavity strongly coupled optical systems operating at room temperatures.

**1. Introduction**

High-index dielectric nanostructures are emerging as alternatives to plasmonic resonators in manipulating light at the nanoscale[1–3]. They possess unique optical properties that are difficult to realize with plasmonic structures. For example, exciting a magnetic dipole mode in a plasmonic structure requires complex geometrical designs [4,5], whereas it is possible to excite multipoles of both electric and magnetic Mie modes in a single dielectric nanodisk [6]. Unlike plasmonic structures, fields can be localized inside the dielectric nanostructures, which allows for light-matter interactions within the structure. However, in most cases, light confinement in dielectric resonators is weaker compared to plasmonic resonators. Therefore, a hybrid design consisting of metal and dielectric nanostructures benefits from the merits of the two systems. By aligning the resonant modes of a dielectric resonator and a plasmonic resonator, the directionality of the scattered light can be better controlled [7–9]. Exciting a plasmon and a Mie mode in a hybrid structure also enhances the local electric field that can further boost the non-linear optical effects from the nanostructures [10].

In the context of mode interactions in photonic/plasmonic structures, there are several coupling regimes based on the coupling strength ($g$) and the damping rates ($\gamma_1, \gamma_2$) of the modes. Broadly, the coupling is classified as weak if $g \ll \gamma_1, \gamma_2$ and as strong, if $g \gg \gamma_1, \gamma_2$ [11]. To attain optical strong coupling, the quality factor of the modes should be large and modal volume should be small [12]. Even though the quality factors of the individual resonators (plasmonic or dielectric) may not be very large, they offer tiny mode volumes by confining light in the subwavelength regimes. Therefore, to reach strong coupling between two resonators it is crucial to excite tightly confined optical modes in them. When the modes interact strongly, hybrid modes are created whose optical characteristics have properties of both the individual modes [12,13]. In this context, resonant coupling of plasmonic and dielectric resonant structures offer unique possibilities to obtain such hybrid modes. Studies of plasmonic-dielectric hybrid systems are still at an early stage [14–16]. Plasmonic nanostructures are very efficient in confining light at subwavelength scales, but performance of dielectric nanostructures in this aspect is poor. But in recent years, optical states such as anapole and bound states in a continuum (BIC) have been demonstrated to show strong confinement of light in dielectric nanostructures and thereby, possibilities of novel light-matter interactions [17,18]. The anapole optical state (hereafter, simply referred to as anapole) can be excited efficiently in a high-index dielectric nanodisk of a specific aspect ratio, where the diameter the disk is large compared to the thickness. Such a nanodisk supports Cartesian electric and toroidal dipoles moments whose far-field radiation patterns are similar. The anapole is generated when the two modes are out-of-phase and destructively interfere in the far-field [17]. This results in a scattering minimum in the far-field and a strong enhancement of light in the near-field.

In this work, we present a metallo-dielectric hybrid metasurface that makes use of an anapole to tightly confine light inside a high index dielectric nanostructure and a surface plasmon polariton (SPP) mode on the metal-dielectric interface to achieve strong coupling between the anapole and the SPP. Specifically, the investigated metallo-dielectric hybrid metasurfaces consists of a high-index dielectric (silicon) nanodisk array on top of a metal layer (aluminium) separated by an oxide (silica) layer. The electromagnetic design and simulations are performed using the finite-difference time-domain (FDTD) method. The physical dimensions of the Si nanodisks are optimized to support the anapole state and the nanodisk array is designed to excite surface plasmon polariton (SPP) at the metal-buffer oxide interface. These parameters are tuned to excite the respective modes at normal incidence of light in the visible-NIR (400-900 nm) wavelength range. The two modes strongly couple at zero detuning of the resonances when the spacer ($SiO_2$) thickness layer is ~200 nm. The avoided crossing of the modes observed in the reflectance spectra and in the spectral profile of light absorption inside the Si nanodisk are signatures of the strong coupling. A vacuum Rabi splitting of up to ~ 129 meV is achievable in the proposed metallo-dielectric hybrid metasurface.

## 2. Metasurface design and optical simulations

The proposed hybrid metasurface consists of silicon (Si) nanodisk arrays in a square lattice on top of a thin (spacer) oxide (SiO2) layer followed by an optically thick aluminum (Al) film on a substrate. The nanodisk dimensions are tuned to support anapole modes and the grating array, in our case a square lattice, is designed to excite SPP at the Al-spacer oxide interface. A schematic representation of the metasurface design is shown in Fig 1a. Although such designs have been demonstrated for strong coupling between localized surface plasmons (LSPs) and SPPs, but not in analogous metallo-dielectric hybrid designs [19,20]. Recently, a similar metasurface design was used to demonstrate beam steering through tunable coupling between Mie-SPP modes [8]. However, this work addresses only electric and magnetic dipoles in the nanodisks and the coupling between these modes with SPP modes is not in strong regime.

The electromagnetic simulations on the presented hybrid metasurface design were performed using a commercial FDTD tool – Lumerical [21]. The refractive index values for Al, SiO2 and Si were taken from Palik database. Total-Field Scattered-Field (TFSF) source was used to calculate the scattering cross-section of an isolated nanodisk. The induced electric fields inside the nanodisk were also collected from the same simulations to estimate multipole contributions to the total scattering cross-section. The multipole decomposition (MPD) of the induced fields was evaluated using an in-house multipole analysis code based on a method developed by Alee et al., [22]. Plane-wave source was used to simulate the total reflectance from the hybrid metasurface. The source was injected along the z-axis. Periodic boundary conditions were used in x and y-normal plane, whereas perfectly matched layer (PML) boundary conditions were used in the z-normal plane.

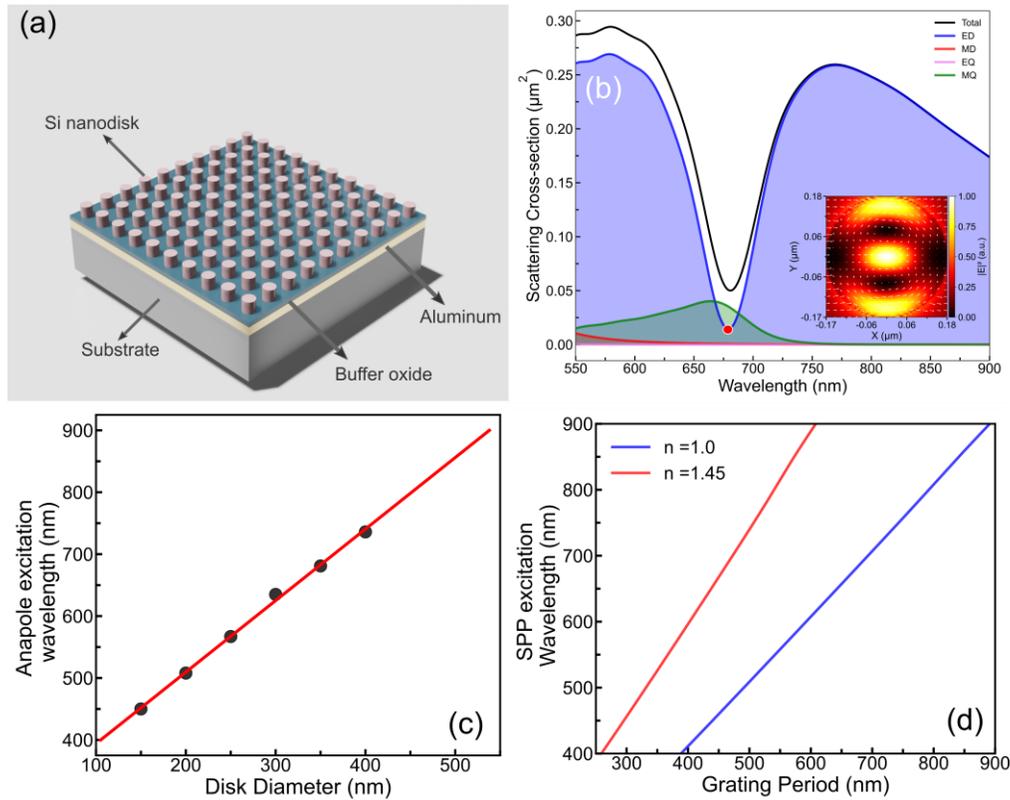

Fig 1. (a) A schematic of metallo-dielectric hybrid metasurface (b) A typical scattering cross-section of a Si nanodisk (height 50 nm and diameter 350 nm) supporting an anapole state. Also shown are the multipole decomposition of the total scattering cross-section. Excitation of the anapole results in a distinctive scattering dip (@ 680 nm) and a strong light confinement within the nanodisk as seen in the cross-sectional field intensity profile. (c) The anapole resonance position in a Si nanodisk can be spanned across Vis-NIR (400-900 nm) wavelengths by changing the nanodisk diameter from 100 to 500 nm. (d) The graph describes the range of grating periods required to tune (1,0) SPP resonance mode in the given wavelength range (400-900 nm), for two different media (n=1 and n=1.45) on top of Al film. Coupling between the modes occurs close to the spectral overlap of the resonances.

Fig 1b shows the scattering cross-section of an isolated silicon disk (diameter D=350 nm, height H=50 nm) supporting an anapole state. The multipole decomposition of the total scattering cross-section shows that the major contribution to the scattering comes from the electric dipole. The influence of the magnetic dipole is minimal in these wavelength ranges due to low aspect ratios (shorter height and larger diameter). The excitation of anapole results in a characteristic dip in the scattering cross-section as seen in Fig1b. Strong localization of the electric fields inside Si nanodisk is evident from the field profiles shown in the inset of Fig1b; the two regions of circulating electric fields that are typical in an anapole state are also illustrated. However, there is a finite light scattering from the magnetic quadrupole close to the anapole resonance position due to which a complete suppression of scattering in the far-field is not possible. The anapole resonance wavelength is tunable (Fig 1c), almost linearly, by changing the nanodisk diameter. For a disk height of 50 nm, increasing the disk diameter from

100 to 500 nm results in a shift of the anapole resonance wavelength by as much as 400 nm, from 400-900 nm. A surface plasmon polariton (SPP) mode is excited at Al-SiO2 interface through grating coupling method. The Si nanodisk square array on top of the spacer oxide serves as a 2D grating necessary to couple incoming light to the SPP mode. For a given period of the 2D grating ($\Lambda$), the SPP excitation wavelength ($\lambda$) is estimated using the SPP dispersion relation and the momentum matching conditions. At normal incidence of light,

$$\Lambda = \lambda \, (i^2 + j^2)^{\frac{1}{2}} \left( \frac{\epsilon_m + \epsilon_d}{\epsilon_m \epsilon_d} \right) \quad , \tag{1}$$

where $\epsilon_m$ and $\epsilon_d$ are permittivities of the metal and the dielectric, respectively, and $i$ and $j$ are the grating orders. Fig 1d describes the shift in the SPP excitation wavelength with changing period of the 2D grating. Here, the SPP is coupled through (1, 0) grating order. The range of grating periods required to excite SPP at the metal-dielectric interface in the wavelength range of 400-900 nm is shown in the figure. Importantly, SPP modes are sensitive to the refractive index of the dielectric material at the metal surface. Therefore, the grating period required to excite SPP at a given wavelength differs based on the refractive index of the dielectric. Fig 1d also shows the relation between the grating period and the SPP excitation wavelength for two different media (air n =1.0, SiO2 n = 1.45). This role of the refractive index has a consequence on the interaction of the anapole and the SPP modes.

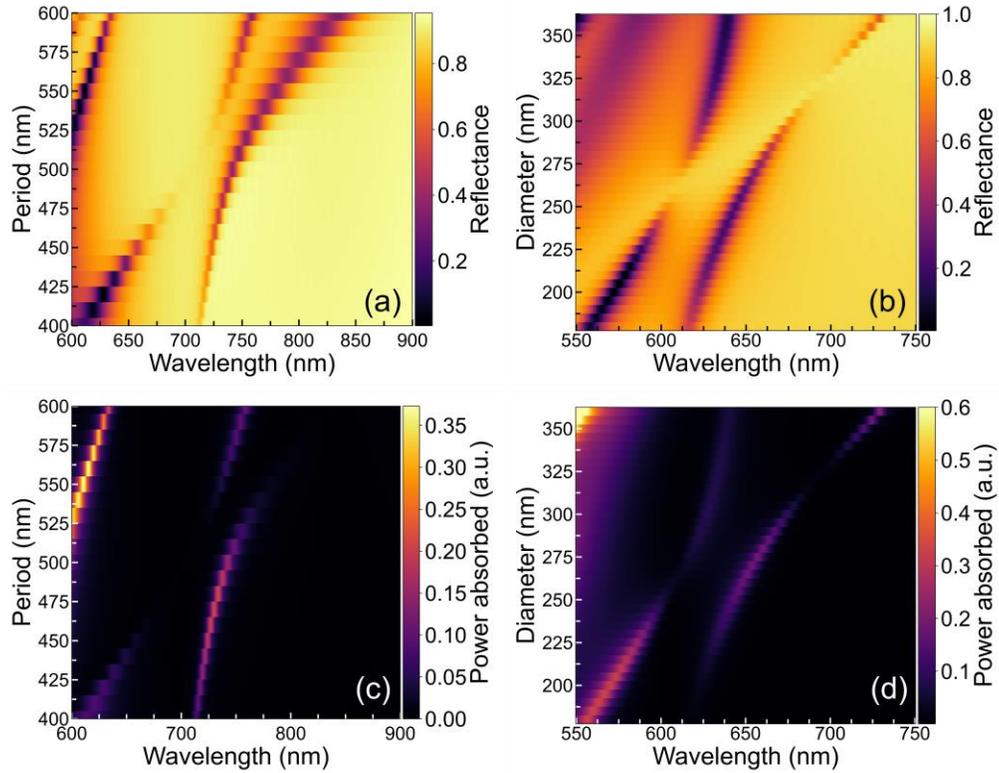

Fig 2. Total reflectance from the hybrid metasurface when (a) the grating period is varied from 400-600 nm to tune the (1,0) SPP mode (at Al-SiO2 interface) and the diameter is kept constant at 350 nm, (b) the diameter is varied from 180-370 nm to tune anapole mode and the grating period is kept constant at 430 nm. Spectral profile of the light absorption within the silicon nanodisk for the case of (c) period sweep and (d) diameter sweep.

## 3. Results and Discussion

As mentioned previously, the SPP resonance is tuned by changing the nanodisk grating period, whereas the anapole state is tuned by varying the diameter of the nanodisk. The two resonances can couple when the nanodisk is in the vicinity of the SPP near-field at the metal-dielectric interface, implying that the nanodisk grating must be placed within the penetration depth of the SPP into the dielectric region. The proximity of the Si nanodisk grating from the Al surface can be varied by changing the spacer oxide layer thickness (and/or the refractive index). In the wavelength range of 400-900 nm, the penetration depth of SPP (at Al-SiO2 interface) into the dielectric region is roughly 125-450 nm. Therefore, the oxide thickness is varied from 25 nm to 200 nm to examine the effect of spatial separation of the modes on their coupling strength. The coupling between the modes were investigated in the far and near-field regimes. In the far-field regime, reflectance from the metasurface is simulated. The anapole and the SPP mode manifest as dips in the reflectance spectra. In the near-field regime, light absorption within the Si nanodisk is calculated. Si absorbs light in the considered wavelength region (400-900 nm). By studying the frequency dependent absorption one can investigate near-field interactions of the modes. The total power absorbed within a material with volume $V$ is given by $\int_V 0.5\omega\epsilon''|E|^2 dV$, where $\omega$ is the angular frequency of the incoming light, $\epsilon''$ is the imaginary part of the permittivity of the material and $E$ is the induced electric field inside the material. A power absorption analysis tool available in Lumerical was used for this purpose. The grating period and the nanodisk diameter were taken as two independent parameters so that only one of the modes is spectrally tuned while the other is kept constant. Fig 2a describes change in the reflectance from the metasurface when the grating period is varied from 400 to 600 nm and the diameter of the nanodisk is kept constant at 350 nm. Here, the thickness of the oxide layer is 200 nm. The (1,0) SPP mode, excited at Al-SiO2 interface, is tuned from 600 to 850 nm. The avoided crossing of the two modes is clearly visible between the spectra corresponding to the grating periods of 480 nm to 530 nm, where a spectral overlap of their resonances is expected. Similarly, avoided crossing between the modes is also observed (Fig 2b) by tuning the diameters of the nanodisk from 180 nm to 360 nm and by keeping the grating period constant at 430 nm. Unlike in a plasmonic resonator-on-metal setup, SPP near-field can penetrate the dielectric resonators of the hybrid metasurface, thereby interact with the modes confined inside the individual nanodisk resonators. The wavelength dependent light absorption profiles are shown in Fig 2c and d for the two parameter sweeps. The avoided crossing of the modes in the absorption profiles is clearly visible confirming the presence of a strong coupling of the anapole and the SPP mode.

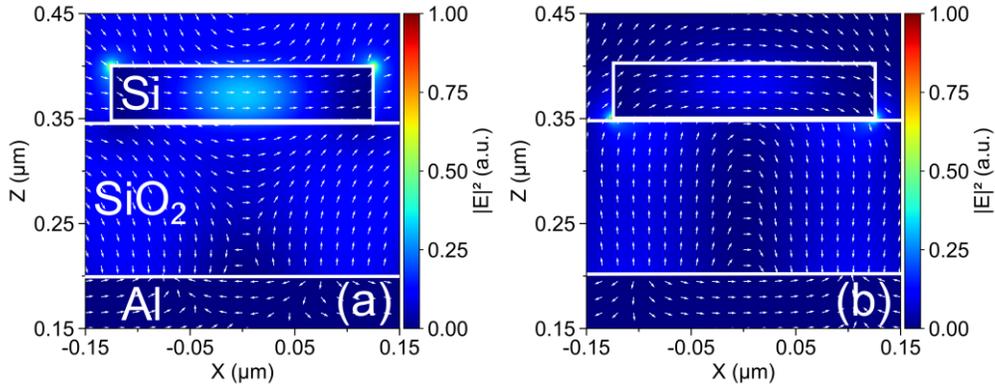

Fig 3. Electric field profiles of a unit cell in the hybrid metasurface with nanodisk diameter of 250 nm and grating period of 430 nm. The profiles display the hybridization of the anapole and the SPP modes at wavelengths (a) 594 nm and (b) 630 nm.

The anapole state for an isolated silicon nanodisk of diameter 350 nm is expected to be around 680 nm (Fig 1a), but we observe that the anapole dip position in the reflectance and power absorptance spectra red shifts with the array period (Fig 2 a, c). This behavior can be attributed to the effect of the lattice on both electric and other multipoles [23,24]. Recently, lattice dependent absorption of anapole modes in GaP nanodisk arrays was reported, in which the shift in the extinction spectra was attributed to the interference of a periodicity dependent lattice mode with the electric and toroidal dipoles [25]. Additionally, there is a resonance shift in the SPP mode due to changing diameter of the nanodisks in the array. This can be attributed to a change in the fill factor of the 2D grating layer with increasing diameter that affects the effective refractive index of the layer. Therefore, the shift in the SPP position with the grating period (Fig 2 a, c) doesn't overlap with the shift in the case of dielectric medium with refractive index of 1.45 (Fig 1d). Another parameter that affects the coupling between the modes is the spacer layer thickness. The total reflectance simulations on the hybrid metasurface with different spacer layer ($SiO_2$) thickness (Supplementary material S1, S2) shows that decreasing the $SiO_2$ layer thickness reduces the coupling strength between the two modes. A clear avoided crossing of the modes starts appearing only for SiO$_2$ layer thicknesses above 150 nm and below 300 nm. The spacer dielectric above the metal surface cannot be treated as a continuous medium with a fixed refractive index. Rather, one must take into account an effective refractive which is determined by the refractive index of the spacer oxide and the nanodisk array (material and spatial arrangement). For example, using zeroth order effective medium theory (EMT), a 2D silicon grating can be converted to a slab waveguide that incorporates both TE and TM effective refractive indices. Such a formulation has been used recently to elucidate the relation between Mie and leaky Bloch mode resonances in an array of scatterers [26].

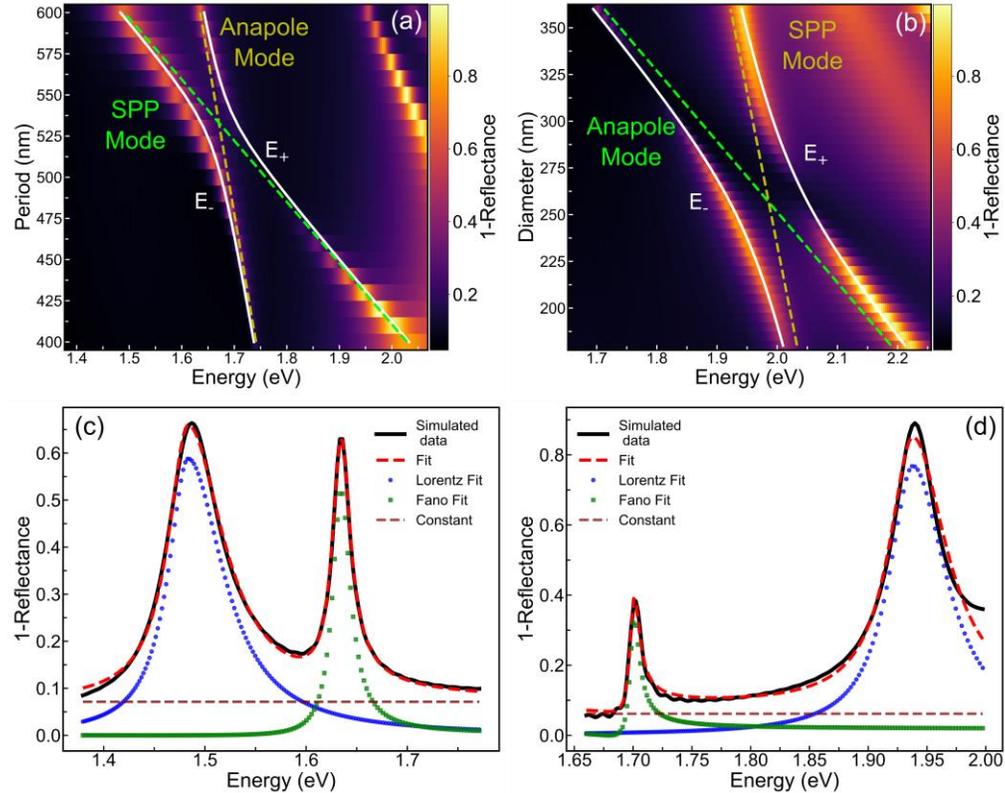

Fig 4. Coupled oscillator model (COM) fit for the normal modes in the extinction spectra (1-Reflectance). (a) The extinction spectra obtained from the grating period sweep (400 – 600 nm) with fixed nanodisk diameter (350 nm). $E_+$ upper branch and $E_-$ lower branch fits are indicated with solid white lines. The green dashed line and the yellow dashed lines show the decoupled SPP and the anapole modes respectively. (b) A similar fit procedure conducted for the case of diameter sweep (180 nm to 360 nm) and a fixed grating period of 430 nm. (c) Fano and Lorentzian fits on the decoupled modes (@ grating period of 600 nm and nanodisk diameter of 350 nm) in period sweep (Fig 4a). and (d) Fano and Lorentzian fits on the decoupled modes (@ nanodisk diameter of 360 nm and grating period of 430 nm) in diameter sweep (Fig 4b). The fit parameters are used in estimating the line width and resonance energy values of the modes.

Fig 3 shows the cross-sectional electric profile of a unit cell in the hybrid metasurface that demonstrates the hybridization of the anapole and the SPP modes. The profile is taken at two wavelength positions, Fig 3a – 594 nm and Fig3b – 630 nm, where the avoided crossing of the modes appears (Fig 2c). The corresponding diameter of the Si nanodisk is 250 nm and the grating period is 430 nm. The field intensity profiles, and the field vector lines indicate that both the anapole and the SPP features exist in the two new hybridized modes. Although, anti-crossing of the modes near zero-detuning is apparent from Fig 2, to ensure that the coupling is indeed in the strong regime, the coupled oscillator model (COM) is implemented. In this model, the energies ($E_+$ upper branch, $E_-$ lower branch) of the hybrid modes for a lossless coupled resonator system is given by [15,27]

$$E_\pm(x) = (E_A(x) + E_{SPP}(x))/2 \pm \sqrt{\left(E_A(x) + E_{SPP}(x)\right)^2/4 + g^2} \qquad (2)$$

Where, $g$ is the coupling strength between the two oscillators, $E_A$ and $E_{SPP}$ are the resonance energies of the decoupled anapole and SPP modes respectively. As the SPP mode is lossy and the anapole mode is excited in a wavelength region (400-800 nm) where silicon has finite absorption, the normal modes are to be considered along with their damping rates. For a system with lossy modes, the condition for strong coupling is given by [12,28]

$$2g > \sqrt{\frac{\gamma_A^2 + \gamma_{SPP}^2}{2}}, \qquad (3)$$

where, $\gamma_A$ and $\gamma_{SPP}$ are the damping rates for the decoupled anapole and SPP resonances respectively. The damping rates of the original states are extracted from the simulated data at a period/diameter where the resonances of the two oscillators are detuned considerably. The anapole is fitted with a Fano line shape of the form [11,29]

$$f(E) = \frac{A(q\Gamma/2 + E - E_0)^2}{(\Gamma/2)^2 + (E - E_0)^2}, \qquad (4)$$

where, $A$ is the amplitude, $q$ the Fano parameter, $\Gamma$ and $E_0$ are the resonance linewidth and energy, respectively. Split Lorentzian lineshape is used to fit the decoupled SPP [12].

Fig 4a and b describe the extinction spectra (1-Reflectance) for the period sweep (400 – 600 nm @ nanodisk diameter of 350 nm) and the diameter sweep (180-360 nm @ grating period of 430 nm), respectively. In the figures, the wavelength ranges (400-900 nm) are expressed in corresponding energy scales (2.25-1.65 eV). The variation of the anapole and the SPP resonance positions with array period is shown on Fig. 4a, dashed lines. The lattice parameters affect the absorption of anapole modes in the nanodisk [25], due to which the anapole dip position shows a small red shift (~ 45 nm) with increasing grating period. Therefore, the fitting parameters to the anapole mode extinction spectrum are extracted from the extinction spectrum of the lattice rather than from the scattering spectrum of an isolated individual nanodisk. The SPP resonance position shows the expected shift with the array period. For the case of period sweep (Fig 4a, in the range 400 – 600 nm), the detuned resonance positions of the anapole and the SPP mode are considered at the grating period of 600 nm with the nanodisk diameter of 350 nm. Whereas, in the case of diameter sweep (Fig 4b), the detuned resonance positions of the anapole and the SPP mode are considered at disk diameter of 360 nm with the grating period of 430 nm. For the above two cases, the Fano and Lorentzian fits for the decoupled anapole and the SPP modes are shown in the Fig 4c and d, respectively. For the case of period sweep (Fig 4c), the anapole linewidth is $\gamma_A$= 21.4 meV and the SPP linewidth is $\gamma_{SPP}$= 65.9 meV and for the case of diameter sweep (Fig 4d), the anapole linewidth is $\gamma_A$ =9.8 meV and the SPP linewidth is $\gamma_{SPP}$=59.0 meV. Equation (1) was fit to the normal modes in the extinction spectra with $g$ (coupling strength) as a single fitting parameter. The fit results in a coupling strength of ~49 meV and ~64.5 meV for the cases of period and diameter sweep respectively. With the calculated damping factors and coupling strengths, Eq.2 is satisfied in both the cases and demonstrates that the system is in the strong coupling regime. The corresponding vacuum Rabi splitting energies are ~98 meV and ~129 meV. These values are comparable with the recent results of vacuum Rabi splitting from strong coupling in Mie resonator based open cavities [1,4,8].

### 4. Conclusion

In summary, we have proposed a metallo-dielectric hybrid metasurface design consisting of high index (Si) nanodisks square array grating on top of a metal film (Al). The 2D grating is separated from the metal surface by a spacer oxide ($SiO_2$) layer. Nanodisks with specific geometries – larger diameter compared (~200 - 350 nm) to the thickness (~50 nm) - support a tightly confined anapole state. The grating array (period~400 -600 nm) of the nanodisks can excite SPP modes at Al-$SiO_2$ interface through grating coupling in the wavelength range of 400- 900 nm. The anapole resonance position can be tuned in a simple way - by changing the diameter of the nanodisk and the SPP resonance can be tuned by changing the period of the 2D grating. FDTD simulations show that a strong coupling of the anapole and the SPP mode is possible at normal incidence of light, when the location of the grating is in the vicinity of the near-field of the excited SPP and at zero detuning of the two modes. The total reflectance from the hybrid metasurface shows avoided crossing of the modes hinting at a strongly coupled system. The avoided crossing is also observed in the near field absorption of the modes inside the Si nanodisk. The coupling strength is sensitive to the spacer ($SiO_2$) thickness, and is stronger for thicknesses between ~100-300 nm. The variation in the coupling strength could be due to changing effective refractive index above the metal-dielectric interface that influences SPP dispersion and field confinement. Coupled oscillator model was implemented to estimate the vacuum Rabi splitting. For an optimized hybrid metasurface design, splitting of up to ~129 meV is possible. The advantages of the investigated hybrid metasurface are that it enables a room temperature open cavity system supporting strongly coupled modes, the design is simple, and fabrication is relatively straightforward using CMOS compatible materials and process technologies. Although the investigations focus on normal incidence of light, interesting possibilities to tune the mode-coupling can be anticipated at off-normal incidence of light, since the anapole states are relatively insensitive to the angle of incidence. Such hybrid metasurfaces with strong mode-coupling could be of high relevance in the fast developing on-chip photonic quantum device technologies.

**Acknowledgements**: The authors acknowledge funding from the Swedish Research Council, grant number: 2019-05321.

**Supplementary material**

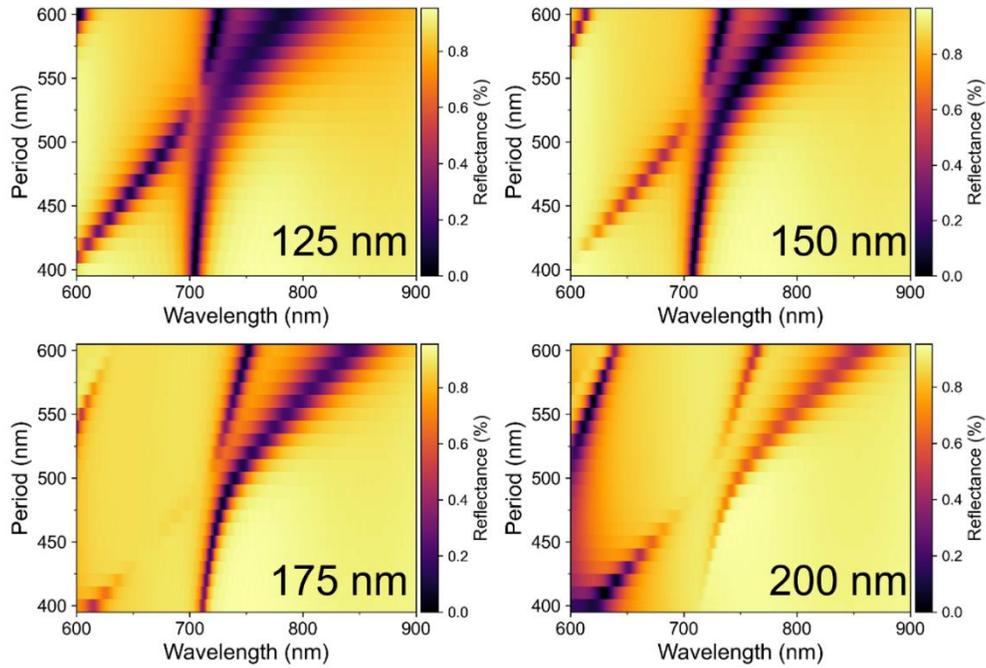

 S1. Reflectance from the hybrid metasurface for different thicknesses of the buffer oxide layer. It is evident that the coupling between the anapole and the SPP is sensitive to the buffer layer thickness. This could be a result of the sensitivity of SPP modes towards the changing effective refractive index of the dielectric layer, above the metal surface, with changing oxide thickness. Importantly, the effective index of the grating layer itself may vary while sweeping the grating period.

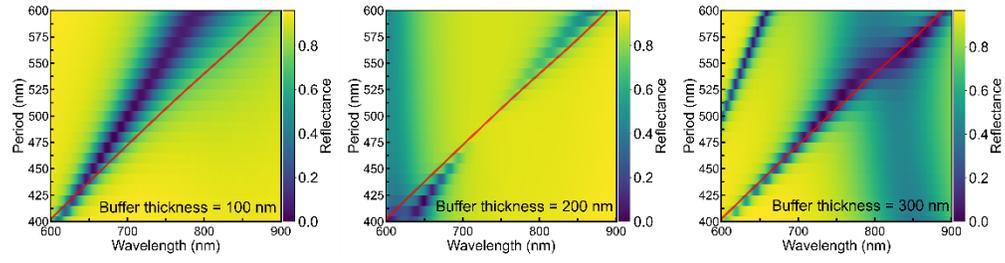

**S2**. Reflectance from a 1D Si grating placed on top of Al surface separated by a buffer oxide of different thickness. The thickness of the Si grating strips is 50 nm and the width is 350 nm. The grating excites only SPP mode at the metal-dielectric interface. The red line indicates shift in the (1, 0) SPP mode with changing grating period, considering that a continuous dielectric medium (SiO2, RI = 1.45) is present above the metal surface.